# The search for decaying Dark Matter

J.W. den Herder (SRON, Netherlands), A. Boyarsky (ETHZ, Switzerland & BITP, Ukraine), O. Ruchayskiy (EPFL, Switzerland)

on behalf of

K.Abazajian (University of Maryland, USA), C. Frenk (ICC, Durham, UK), S.Hansen (NBI, Denmark), P. Jonker (SRON, Netherlands), C. Kouveliotou (NASA, USA), J.Lesgourgues (CERN, Switzerland & CNRS, France), A. Neronov (ISDC, Switzerland), T. Ohashi (MTU, Japan), F. Paerels (University of Columbia, USA), S. Paltani (ISDC, Switzerland), L. Piro (INAF-IASF/Rome Italy), M. Pohl (Université de Genève, Switzerland), M. Shaposhnikov (EPFL, Switzerland), J.Silk (Oxford, UK), J.Valle (Valencia, Spain)

## Abstract

We propose an X-ray mission called Xenia to search for decaying superweakly interacting Dark Matter particles (super-WIMP) with a mass in the keV range. The mission and its observation plan are capable of providing a major break through in our understanding of the nature of Dark Matter (DM). It will confirm, or reject, predictions of a number of particle physics models by increasing the sensitivity of the search for decaying DM by about two orders of magnitude through a wide-field imaging X-ray spectrometer in combination with a dedicated observation program.

The proposed mission will provide unique limits on the mixing angle and mass of neutral leptons, right handed partners of neutrinos, which are important Dark Matter candidates. The existence of these particles is strongly motivated by observed neutrino flavor oscillations and the problem of baryon asymmetry of the Universe.

In super-WIMP models, the details of the formation of the cosmic web are different from those of ΛCDM. The proposed mission will, in addition to the search for decaying Dark Matter, provide crucial insight into the nature of DM by studying the structure of the "cosmic web". This will be done by searching for missing baryons in emission, and by using gamma-ray bursts as backlight to observe the warm-hot intergalactic media in absorption.

# 1. Dark Matter in the Universe

The nature of Dark Matter is one of the most intriguing problems of modern physics. In particular, it is crucial for building a fundamental theory of particle physics. Indeed, while the Standard Model of elementary particles (SM) describes with great precision the particle physics data, it *does not provide* a DM candidate. In fact, the dominant fraction of DM cannot be made out of baryons, as it would be impossible to generate such an amount of baryonic matter in the scenario of the otherwise successful Big Bang nucleosynthesis theory. The only viable non-baryonic candidate, the neutrino, would lead to the top-down scenario of structure formation and is ruled out from the cosmological observations of the large scale structure. Therefore, the DM particle hypothesis implies an *extension of the SM*. By constraining properties of DM particles, one can differentiate among different extensions of the Standard Model and shed light onto the fundamental properties of matter.

## 1.1 WIMP Dark Matter candidates

Maybe the most popular DM candidates are *weakly interacting massive particles* (WIMPs). These stable particles interact with the SM with roughly electroweak strength. The interest in this class of candidates is due to their potential relation to the electroweak symmetry breaking, which will be tested at the Large Hadron Collider in CERN. To give a correct DM abundance, these particles should have a mass from ~1 GeV to ~$10^3$ GeV. The WIMP searches are important scientific goals of many experiments. Dozens of *dedicated* laboratory experiments (e.g. CDMS, USA; DAMA/LIBRA & XENON10, Italy; ArDM, CERN; KIMS, Korea) are conducted to detect the interaction of WIMPs in the Galaxy's DM halo with laboratory nucleons. One of the main scientific objectives of the recently launched *Fermi* mission is to search for gamma-rays from WIMP DM annihilation and thus explore "signs of new laws of physics and what composes the mysterious Dark Matter". WIMPs appear in many extensions of the Standard Model (super symmetric models, models with extra dimensions, etc.).

## 1.2 Super-WIMP DM candidates

Another large class of DM candidates are *superweakly* interacting particles (so called super-WIMPs), i.e. particles, whose interaction strength with the SM particles is much more feeble than the weak one. Super-WIMPs appear in many particle physics models: extensions of the SM by right-handed neutrinos [6], [7], [36], Supersymmetric theories [21], [20], [51], models with extra dimensions and string-motivated models [22]. Compared to WIMP DM candidates the properties of super-WIMP DM candidates are different in two crucial ways:

- their mass can be as low as a ~ 0.5 keV;
- they can *decay* into the SM particles.

Indeed, all the above-listed models possess a 2-body radiative decay channel: DM → γ + ν, γ + γ, producing a photon with energy E$_\gamma$ = M$_{DM}$/2. Although the super-WIMP DM has a cosmologically long lifetime and decays rarely, it can produce a monochromatic decay line in the spectra of DM-dominated objects. Such line, if detected, will uniquely define the properties of these super-WIMP DM candidates. Searching for such a line provides a way for the detection of DM particles [1],[3],[17],[24].

### Sterile Neutrinos

One particular super-WIMP DM candidate recently attracted a lot of attention [19]. In the SM all fermions (apart from neutrinos) have both left- and right-handed components. The structure of the SM dictates that right-handed partners of neutrinos would not be charged with respect to the Standard Model interactions and interact with other matter only via the mixing with the usual (left-handed) neutrinos (that is why they are often called *sterile neutrinos*). Sterile neutrinos with a mass in the keV range could be produced in the Early Universe with the correct abundance to account for all the DM [23],[46],[24],[3],[6],[7]. Thus it provides a viable DM candidate. Via its mixing with the SM neutrinos, the sterile neutrino can decay into a neutrino and a photon.

*Neutrino oscillations and baryon asymmetry.*

An extension of the Standard Model with three right-handed neutrinos (called **νMSM**, neutrino extended Minimal Standard model [6],[7],[19]) is able to explain several other observed phenomena beyond the SM. Apart from the absence of a DM candidate, the SM fails to explain the observed phenomena of neutrino oscillations – the transition between neutrinos of different flavors. The only explanation of these phenomena is the existence of mass of active neutrinos. The most natural way to provide this mass is by adding the sterile neutrinos. Moreover, as demonstrated by [6],[7], the parameters of the right-handed neutrinos can be chosen in such a way that such a model resolves another problem of the SM – it explains the excess of baryons over antibaryons in the Universe (the baryon asymmetry).

*The νMSM provides an explanation of these observed phenomena beyond the SM and the DM candidate within one consistent framework. This makes it important to test this model experimentally.*

The masses of all new particles in νMSM are entirely within a reach of modern accelerators, which makes this theory potentially testable. Searches for sterile neutrinos can be conducted in particle physics experiments [26]. The search of these particles at accelerators is included in the program of the future experiment NA62 at CERN [40] and in the proposals NuSOnG [5] and HiResMν [39] at FNAL. There is a possibility to improve the existing bounds at the MINERνA experiment [53] and to make dedicated experiments at CERN PS and SPS accelerators. The lightest sterile neutrino can be searched for in precision experiments on double β decay [8]. The best way to search for DM sterile neutrinos is via astrophysical observations discussed in this proposal.

*Other models of sterile neutrino DM*

The νMSM is a minimal model, containing neutrino oscillations and DM sterile neutrinos. Other extensions of the Standard Model, motivated by additional particle physics and cosmological phenomena, also predict production of keV mass sterile neutrino DM. In these models these particles can be produced e.g. in decays of inflaton [48] gauge-singlet scalar field [33], [42], [11] or new weakly interacting bosons [32] ,[57].

**Light gravitinos**

Another interesting example of superweakly interacting decaying DM candidate is R-parity violating gravitino. Among super symmetric extensions of the SM there is a class of models [28] in which the gravitino mass is in the keV range. The gravitino is then the lightest super symmetric particle and is very long-lived (if R-parity is mildly violated). If the maximal temperature reached by the primordial plasma is high enough, then gravitinos are produced in superpartner decays and scatterings. Otherwise, they can be produced by the decay of heavy relics or moduli fields. From the particle physics point of view, the models with the keV mass gravitinos can be either confirmed or ruled out at the LHC [27].

**1.3 Cosmological and astrophysical consequences of the keV-mass super-WIMPs**

The super-WIMP DM particles with their masses in the keV range can have a number of interesting astrophysical and cosmological implications.

*Structure formation in the Universe*

The structure formation in the Universe is one of the main pieces of evidence for the existence of DM. The nature of DM leaves its imprints on the currently observed structures and the way they formed. If DM particles are light, and are relativistical in the early Universe, they stream out of the overdense regions and erase primordial density fluctuations below a certain scale. The observations of the galaxy correlation function (LSS) together with the measurements of Cosmic Microwave Background anisotropies rule out the possibility that a significant fraction of DM particles remained relativistic up to the matter-dominated epoch (which is the case for the SM neutrinos). There is, however, a vast gap between such "hot" DM models and WIMPs. DM particles, that are produced relativistically and become non-relativistic during the radiation-dominated epoch, are compatible with the limits from CMB and LSS surveys as well as the Λ-Cold DM "concordance" model. However at roughly galactic scales (below ~Mpc) they possess different clustering properties. Such DM models are often called *warm dark matter* models (WDM) [10].

Being warm DM, super-WIMPs with keV mass modify structure formation and in this way affect the way the first stars are formed. They also influence the reionization of the Universe [47],[41],[25],[49]. Moreover, the photons from decaying DM with the mass in keV range can speed up the gas cooling [9],[49],[50].

A DM in the form of sterile neutrino has a number of additional astrophysical consequences. It affects the physics of supernova explosion [31]. In particular the asymmetric emission of sterile neutrinos may explain large velocities of pulsars [35],[33].

### 1.4 Strategy of search for decaying DM

In the case of the decaying super-WIMPs the astrophysical search becomes very promising. First of all, a positive result would be conclusive, as the DM origin of any ``"suspicious" line can be unambiguously checked. Indeed, the *decay* signal is proportional to the DM density $\rho_{DM}$, integrated along the line of sight (l.o.s.): $\int_{l.o.s.} \rho_{DM}\, dr$ as opposed to the $\int_{l.o.s.} \rho_{DM}^2\, dr$ in the case of annihilating DM. As a result a vast variety of astrophysical objects of different nature will produce a comparable decay signal [18]. Therefore one has *(a)* the freedom of choosing the observational targets, avoiding complicated astrophysical backgrounds and *(b)* if a candidate line is found, its surface brightness profile can be measured (as it does not decay quickly as function of the distance from the centers of the objects it can can be distinguished from astrophysical lines that usually decay in the outskirts). This allows to distinguish the decaying DM line from an emission line from the hot gas in the astrophysical object. Therefore the astrophysical search for the decaying DM is *an unique type of a direct detection experiment*.

It should be stressed that the strategy and feasibility of laboratory searches of super-weakly interacting massive particles is highly sensitive to the values of their masses and interaction strength. The astrophysical searches allow to probe and restrict simultaneously a wide range of parameters. This permits us to constrain the parameter space and to identify the appropriate strategy for the search at ground-based experiments.

## 2. Existing constraints

In this section we describe existing restrictions on super-WIMP DM models.

**Universal lower bound**

Any fermionic DM should satisfy the universal Tremaine-Gunn bound and its mass should be above 300-500 eV [52]. This means, that the lowest energy range in which one can search for the decaying DM is the X-ray.

**X-ray constraints**

Within the last several years an extensive search for a DM decay signal was conducted in the keV-MeV range using archive data of *XMM-Newton* [12],[17],[18],[55] Chandra [4],[44], INTEGRAL [16],[56] and *Suzaku* missions [37]. It did not reveal any "candidate" line in the energy range from ~ 0.5 keV to above 10 MeV. The current exclusion region is shown on figure 1.

**Bounds from structure formation**

In the case of particles whose primordial velocity distribution is close to the thermal one, the *Lyman-α forest method* allows to put an upper bound on their average velocity that can be converted to a lower bound on the mass in a model-specific way. The Lyman-α forest method is a powerful cosmological tool to probe the formation of structures of (sub)Mpc scales. Under the assumption that the distribution of the the neutral hydrogen traces the distribution of DM, one can reconstruct the power spectrum of density fluctuations at redshifts z~ 2-5 and sub-Mpc scales from the statistics of absorption features due to neutral hydrogen in the spectra of distant quasars. At these redshifts the evolution of small scale structures has already entered into the non-linear stage and large suites of numerical simulations are required to model it. Apart from these difficulties, the astrophysics of the intergalactic medium, entering the Lyman-α analysis is complicated, and not yet fully understood. Thus the Lyman-α method should be taken with care and additional methods should be employed.

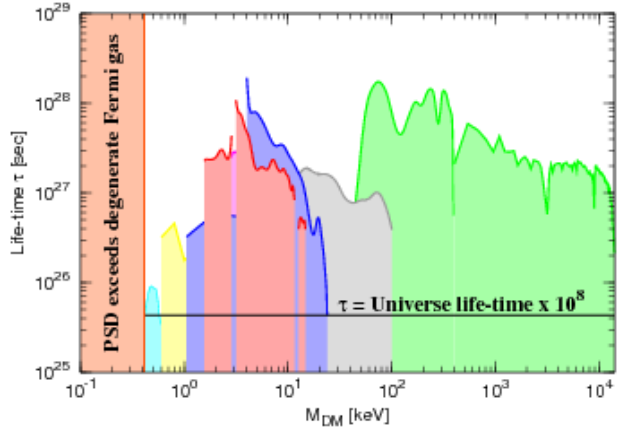

*Figure 1 The lower bound on the lifetime of decaying DM: combined exclusion plot from X-ray observation. The vertical shaded region marks the universal Tremaine-Gunn lower mass bound [52]. The constraints are from Suzaku, XMM-Newton, Chandra, HEAO-1, INTEGRAL (SPI) as well as a prototype spectrometer of McCammon ([38]in [12]).*

In the case of sterile neutrino DM with a thermal spectrum based on the combined analysis of the WMAP5 and SDSS Lyman-α data, it leads to a lower bound on the DM mass of 8 keV [14] revising previous bounds of [29],[2],[45],[54]. However, due to their feeble interaction strength, the super-WIMP DM candidates are often produced out of thermal equilibrium with a non-thermal spectrum (such models may often resemble a mixture of cold and warm DM components). In this case the Lyman-α analysis depends on several parameters of DM models. This makes the mass bounds weaker. For example, it was shown recently [15] that a sterile neutrino DM with the mass as low as 1-2 keV is compatible with all existing bounds including those from the Ly-α method.

**Combined astrophysical constraints**

Figure 2 shows the constraints on the mass and the *mixing angle* $\sin^2(2\theta)$, characterizing the interaction strength between a sterile and an ordinary neutrino. The region between two black thick lines marks the range of parameters for which the sterile neutrinos can be created within the framework of the νMSM with their abundance correctly reproducing the measured density of DM. The colored upper right corner corresponds to a parameter region, excluded from the previous searches for the DM decay line from X-ray observations by Suzaku, XMM-Newton, Chandra, HEAO-1 and INTEGRAL observatories. Finally, a lower limit on the mass of DM sterile neutrino $M_s$ ≥ 1-2 keV comes from the analysis of the Lyman-α forest data in the νMSM model [15].

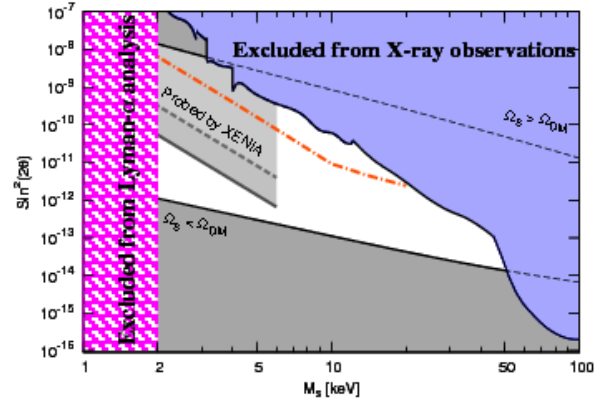

*Figure 2. The allowed region on the mixing angle θ of the DM sterile neutrino with mass $M_s$ (the unshaded region). The red dashed-dotted line shows the predicted improvement of the sensitivity from a 1 Msec observation of a DM dominated object with XMM-Newton's EPIC camera. The light gray region marked "probed by Xenia" shows the improvement with 1 Msec observations of dwarf spheroidal galaxies (dashed) and with the combined analysis of all observations in 2 years. The upper right shaded corner is excluded from previous Suzaku, XMM, Chandra, HEAO-1 and INTEGRAL observations.*

An important feature of figure 2 is that the admitted region is surrounded by different constraints in all directions, telling that the hypothesis of sterile neutrino as a DM candidate is experimentally testable. The range of DM masses, probed by the Xenia mission, is the most interesting from the point of view of the impact on astrophysical phenomena, as well as on the structure formation. It is also the most promising parameter region for ground-based laboratory searches of super-weakly interacting massive particles.

## 3. Additional science

Not described in this white paper is the additional scientific return of the proposed mission. These are described in detail in [43],[30] but some key elements are listed here:

- the Xenia mission will enable mapping and characterizing the cosmic web (missing baryons) up to z~1 using the Oxygen emission lines (O VII and O VIII) down to overdensities of ~ 100 (a factor 30 below the current detection limits);
- trace the evolution and physics of clusters out to their formation epoch (z>1);
- study the evolution of massive star formation using Gamma-Ray-Bursts to trace their explosions back to the early epochs of the Universe (z > 6);
- measure the metals in host galaxies of GRBs and the explosive enrichment in their close environment out to z > 6;
- detect electro-magnetic counter parts for the gravitational wave detections by LISA. The FoV and capability to monitor these areas to separate the real turn-on from variable sources in the FoV (AGNs) will be an important asset of the mission.

## 4. Instrument

Key observational parameters which determine the sensitivity of the proposed instrument are the energy resolution, the effective area and the field of view over a broad energy range. Clearly one likes to maximize all parameters but unfortunately detector technology and X-ray optics have their limitations. For an imaging camera with an energy resolution of R > 500 one requires cryogenic detectors which have a limited size (maximum few x few $cm^2$). For such small detector a large Field-of-View (FoV) corresponds to a short focal length. Unfortunately X-rays only reflect at grazing incidence which limits the size of the effective area. Employing novel four-fold reflections and accepting a limited energy range, it is, nevertheless, possible to define an instrument which can place strong constraints on the properties of sterile neutrinos. This instrument is shown in Figure 3 and its key parameters are given in Table 1. Some further trade-off between the field of view and energy range can be made to optimize the instrument for the search of decaying DM signals.

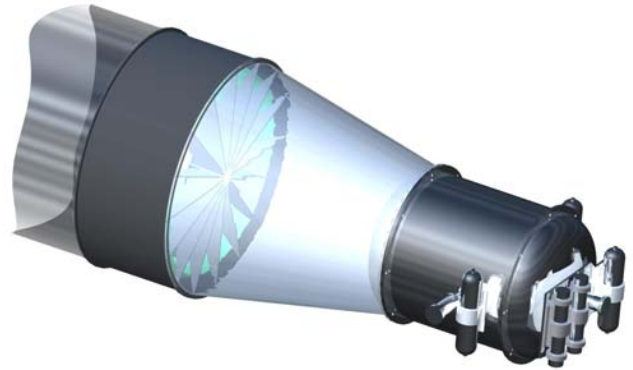

*Figure 3 Schematic drawing of the instrument with the X-ray optics (left) and the dewar containing the cryogenic spectrometer (right). X-rays enter from the left.*

*Table 1 Key instrument parameters (@1 keV)*

| Parameter | Requirement | Goal |
| --- | --- | --- |
| Resolution [eV] | 2.5 | 1 |
| Field of view [$deg^2$] | 0.9 x0.9 | 1.0 x 1.0 |
| Energy range [keV] | 0.2 – 2.2 | 0.1 – 3.0 |
| Effective area [$cm^2$] | 1000 | 1300 |
| Grasp [$cm^2 deg^2$] | 400 | 500 |
| Angular resolution [arcmin] | 4 | 2.5 |

Currently the detector technology is already at Technology Readiness Level (TRL) ≥ 4 with an active program to reach TRL 5-6 in 2011 as the same detector has been selected as baseline for the International X-ray Observatory (IXO). The first segments of the optical reflectors have been produced and have demonstrated an encouraging optical performance of ~ 4 arcmin which is consistent with the requirements. The estimated mass and power of the instrument (including 30% margins) are 575 kg and 1092 W.

## 5. DM search with X-ray missions

**Sensitivity**

Because of its cosmologically long lifetime, the width of the DM decay line is determined entirely by the Doppler broadening due to the virial motion: $\Delta E/E_\gamma$ ranges from ~ $10^{-4}$ for a typical dwarf spheroidal galaxy to ~ $10^{-2}$ for a galaxy cluster. Therefore, the spectral resolution is the crucial parameter in the search of the DM decay line. In the absence of a clearly detectable line, the sensitivity to exclude a decay line against an astrophysical background scales as $\sim 1/\sqrt{(\Delta E)}$. However, drastic improvement of the spectral resolution, as compared to the current X-ray missions may lead to the situation when a strong line is detected in a practically background free regime. The sensitivity towards the detection of such a strong line scales as $\sim 1/(\Delta E)$ [12]. The statistics of both the DM line and background are proportional to the size of the FoV ($\Omega_{fov}$) and effective collecting area, $A_{eff}$ and therefore the sensitivity for the DM line detection scales as $\sqrt{(\Omega_{fov} \times A_{eff})}$. This product is usually called grasp.

Figure 4 compares the sensitivities of existing and planned missions for the detection of DM decay line from the nearby dwarf galaxies or galaxy cluster of the angular size of ~$1^o$ in the parameter space relevant for the diffuse DM line detection: Energy resolution versus grasp. The sensitivity of *XMM-Newton* EPIC camera is taken as a reference. Solid lines indicate improvement of the sensitivity by factors of 1, 10 and 100 (the top left is the most sensitive). The dashed gray lines show the improvement of the sensitivity towards the detection of a *strong line* (in an effectively background free regime). The turnover of the lines at the resolving power $\Delta E/E \sim 10^{-2}$ in figure 4.b is related to the natural width of the line in the galaxy cluster.

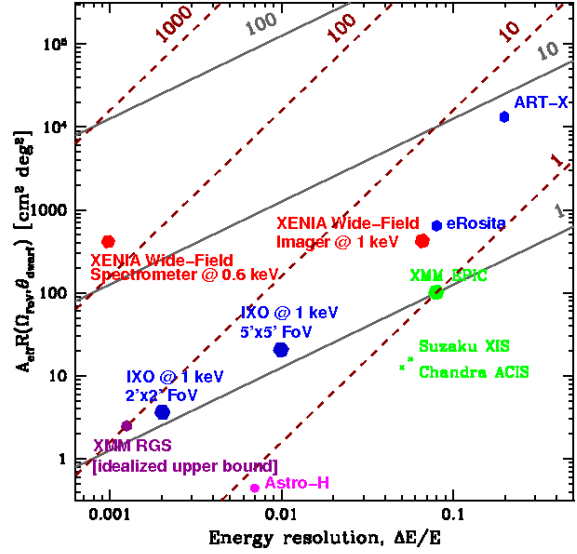

*Figure 4.a Comparison of sensitivities of existing and proposed/planned X-ray missions for the detection of the DM decay line from a nearby dwarf spheroidal galaxy of the angular size of $1^o$.*

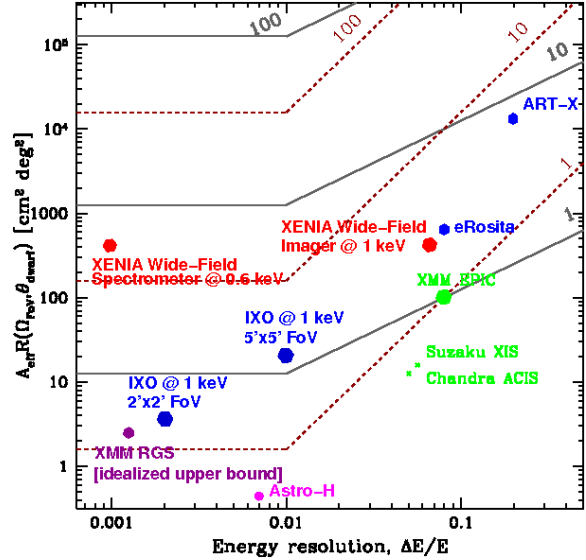

*Figure 4.b Similar to figure 4.a with the exception that the sensitivity is shown for a galaxy cluster.*

**Other X-ray missions**

We see that wide-field spectrometer on Xenia is definitely the best instrument among existing and proposed X-ray missions for the sterile neutrino DM search. The low energy micro-calorimeter detector has the best GRASP among all detectors with high resolving power ΔE/E ~ 0.1 %. On the other hand, among the missions with comparably large grasp (such as eRosita for the planned Russian Spektr-RG mission) the Xenia spectrometer has a much better spectral resolution.

## 6. Observational Strategy

As shown in [17] the signal from almost all nearby objects (dwarf galaxies, Milky Way, large elliptic galaxies, galaxy clusters) provide a comparable (within an order of magnitude) DM decay signal. Therefore, observations of any astrophysical object where the underlying spectrum can be described by a convincing physical model, is well suited for the decaying DM. This strategy can be matched extremely well with an **additional science program** (see section 3)**.** Observations of dwarf spheroidal satellites of our Galaxy are expected to provide the strongest restrictions as these objects have smaller velocity dispersion and thus Doppler broadening compared to large galaxies or galaxy clusters and they are very dark in X-ray, thus optimizing a signal-to-noise ratio. Other types of objects with large DM overdensity will be observed as well (see table 1).

*Table 2. Tentative observing program for the first 2 years (70% observing efficiency, half a year commissioning). The additional astrophysical science program will include extended sources with low surface brightness (cluster outskirts, SNRs, WHIM filaments).*

| Targets | # | Msec |
|---|---|---|
| Dwarf galaxies (Ursa Minor, Draco, Sculptor) | 3 | 12 |
| Milky way | | 4 |
| Andromeda galaxy | 1 | 8 |
| Clusters of galaxies (Coma) | 3 | 12 |

An important aspect in the search for decaying DM is the variable foreground due to our own galaxy. By a dedicated observational strategy this will be approached: (a) performing long observations (typically 1 Ms), (b) on-source and off-source pointings and (c) returning to promising sources more than once. Combining these measurements we typically require 4 Ms per source.

## 7. Space mission

The required instrument has been proposed as core payload for various missions. The *International X-ray Observatory* has a smaller FoV as has been illustrated in figure 4 and observing times of 4 Ms are prohibitively long. The proposed *Xenia* mission or a reduced version of this mission (*New Explorer of the Cosmic Web*) have a cryogenic imaging spectrometer with a large Field-of-View as core payload and are therefore best suited for these measurements.

**Xenia**

The capability of X-ray observations to search for decaying Dark Matter has been recognized and was one of the auxiliary science topics discussed in the EDGE proposal for the Cosmic Vision program of ESA. This EDGE proposal evolved in the Xenia mission, which is currently proposed to the US decadal review as a mission to study *the cosmo-chemical evolution of the Universe*. It includes the required wide-field high spectral resolution imager, needed to detect, or provide limits on parameters of decaying DM. The second imaging instrument on Xenia is a High Angular Resolution Imager using polynomial optics and a CCD camera. This improves the identification and rejection of point source contributions to the larger pixels (4 arcmin) of the cryogenic imaging spectrometer. The third instrument is a transient event Detector with a typical FoV of 2.8 sr. Similar to the USA SWIFT satellite this will autonomously trigger the repointing of the satellite after the detection of an X-ray burst.

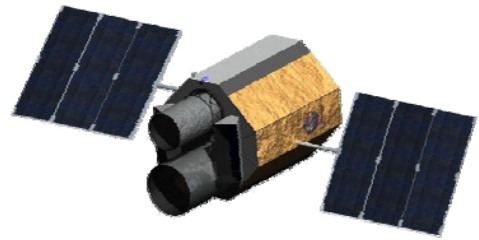

*Figure 5 Xenia after deployment of the solar panels*

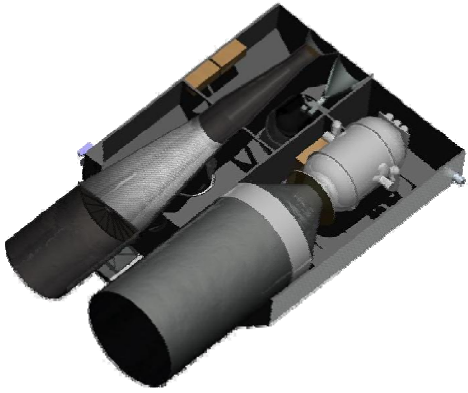

*Figure 6 Xenia showing clearly the cryogenic imaging spectrometer (bottom) and the high-angular resolution imager (top)*

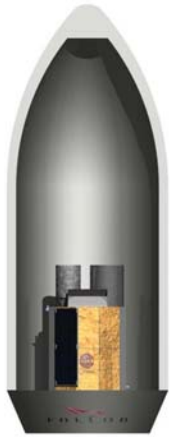

*Figure 7 Xenia in the Falcon 9 fairing*

*Table 3 Key parameters for the Xenia mission*

| parameter | |
|---|---|
| payload | - Cryogenic imaging spectrometer (see text)<br>- High angular resolution imager (CCD camera, < 15 arcsec)<br>- Transient Event Detector (coded mask, CZT) |
| Orbit | LEO, 600 km, inclination < 10$^o$ |
| Mass | 2637 kg (including 30% margin) |
| Power | 2027 W End of Life |
| Slew | Very rapid, 60$^o$ in 60 sec |
| Launcher | Falcon 9, possibly Vega |

**New Explorer of the Cosmic Web**

An attractive reduced implementation has been defined but no detailed assessment studies were carried out. In this concept only the cryogenic imaging spectrometer, in combination with a simplified monitor of transient sources, is selected as payload. This is illustrated in figure 8 and will reduce the mass and power (and thus cost) to < 1600 kg and < 1200 W. Such satellite could, for example, be launched with the European VEGA launcher. By excluding the monitoring and fast repointing capability one could implement this instrument on an even smaller satellite. In principle, one could even consider to drop the optics and use only a simple collimated detector (increases the energy range but decreases the area). However, the scientific return in other fields of astrophysics, fully justify the extra mass, power and complexity, needed for the autonomous fast repointing and focusing optics.

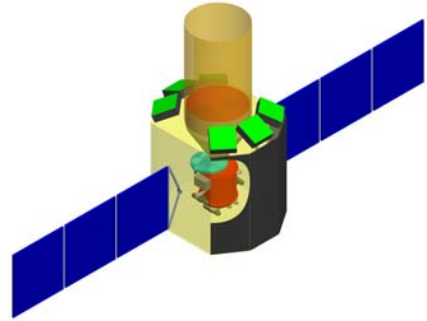

*Figure 8 schematics of the simplified mission with the Cryogenic Imaging Spectrometer and the Transient Event Detector (6 green detectors)*